\documentclass[aps,prb,twocolumn,showpacs]{revtex4-1}

\usepackage{amssymb,amsmath}
\usepackage{graphicx}
\usepackage{enumerate}

\usepackage{balance}
\usepackage{float}
\usepackage{color}

\begin{document}
	
	\title{Charging of highly resistive granular metal films}
	\author{M. F. Orihuela}
	\author{M. Ortu\~no}
	\author{A. M. Somoza}
		\author{J. Colchero}
		\author{E. Palacios-Lid\'on}
		\email{Corresponding author: elisapl@um.es}
	\affiliation{Departamento de F\'isica -- CIOyN, Universidad de Murcia, Murcia 30.071, Spain}
	\author{T. Grenet}
	\author{J. Delahaye}	
	\affiliation{Institut N\'eel, CNRS and Universit\'e Joseph Fourier, B.P. 166, F-38042 Grenoble C\'edex 9, France}
	\date{\today}

\begin{abstract}

We have used the Scanning Kelvin probe microscopy technique to monitor the charging process of highly resistive granular thin films.
The sample is connected to two leads and is separated by an insulator layer from a gate electrode. When a gate voltage is applied,  charges enter from the leads and rearrange across the sample. 
We find very slow processes with characteristic charging times exponentially distributed over a wide range of values, resulting in a logarithmic relaxation to equilibrium.
After the gate voltage has been switched off, the system again relaxes logarithmically slowly to the new equilibrium.
The results cannot be explained with diffusion models, but most of them can be understood with a hopping percolation model, in which the localization length is shorter than the typical site separation.
The technique is very promising for the study of slow phenomena in highly resistive systems and will be able to estimate the conductance of these systems when direct macroscopic measurement techniques are not sensitive enough.
\end{abstract}

\maketitle

\section{Introduction}

During the last decades, slow conductance relaxation has been studied in many disordered insulators and, in particular, in granular metals by means of field effect measurements.
After a quench at low temperatures, a change in the gate voltage is accompanied by a sudden increase in the conductivity, which subsequently slowly decreases with a roughly logarithmic dependence on time \cite{VaOv02,Grenet2007,Ovadyahu2007,HaEi12}.
Memory effects and aging are often seen in the same type of experiments \cite{Grenet2010}.
All these glassy effects have been interpreted in different ways, but there is a growing tendency to explain them in terms of electron glasses, i.e., systems with states localized by the disorder and long-range Coulomb interactions between carriers \cite{Pollak06,POF13,Amir15}.

These glassy effects were first limited to low temperature $T$, but they have recently been observed at room temperature in discontinuous Au \cite{Eisenbach2016}, amorphous NbSi \cite{Delahaye2014} and granular Al films \cite{granular}.
In addition, high resolution techniques such as scanning force microscopy (SFM) open the possibility of studying these systems at  room temperature  allowing a detailed real space analysis of the problem.
Some of us have applied the SFM technique to study slow relaxation in the surface potential of conducting polymers \cite{Ortuno2016}. We  observed  logarithmic relaxation over four decades of time, as well as full aging in terms of the time of application of the gate voltage. The technique proved appropriate to monitor slow relaxation at the nanoscale.

The use of local probe techniques, such as scanning Kelvin probe microscopy (SKPM), presents two advantages as compared with the conductance measurements performed so far: i) it allows a study of the phenomena at the nanometer scale, which can shed some light on the mesoscopic effects at work in the conductance relaxations, previously observed in indium oxide \cite{Orlyanchik} and granular Al films  \cite{Delahaye2008}, but not fully understood; ii) samples with larger resistances can be measured, which is interesting because we know that in the “electron glass” experiments, the larger the resistance, the larger the conductance relaxations (in \%\ of conductance).

In this problem the first question to answer is the mechanism of charge injection in a strongly disordered insulator. It is an interesting general question and has rarely been addressed. The most thorough study on this subject was the work by  Adkins' group \cite{adkins,adkins84}. 
Analysing non local electrical measurements, they found that inhomogeneities play a crucial role in the charging process and that the scale of inhomogeneities grows as the proportion of metal increases and  grains coalesce.
They also found strong evidence for the percolative nature of both transport and the charging process.
Finally, they monitored the total charge accumulated in the sample as a function of time and observed that the corresponding time dependence was clearly broader than the predictions of diffusion models.
In discontinuous metal films and in many other strongly insulating systems, the basic assumption of the conduction models is the exponential-like distribution of the microscopic hopping rates \cite{POF13}. Its implications in macroscopic transport properties have been systematically confirmed. It gives rise, for example, to the widely observed variable-range hopping \cite{TsEf02,SoOr06}, but no direct microscopic evidence have been produced so far.

In this paper, we study with the SFM technique the time evolution and the spatial dependence of the surface potential of granular Al thin films when a gate voltage is suddenly applied and later  on switched off, after a  relatively long period of time. 
We measure samples that are qualitatively similar to those previously studied by some of us in the context of slow relaxation in the conductivity, but that have been grown  with the appropriate resistance for this experiment, that is, with a resistance high enough to ensure that the charging processes take place in the scale  from milliseconds to hours.
As our experimental results are difficult to understand with a model of homogeneous conductivity, we have constructed a more general network resistance-capacitance model able to handle strong inhomogeneities  \cite{Ortuno:2015}.

In the next section we describe the SFM technique employed and how the samples have been grown. In section III we present the experimental results, first for the charging experiments and later for the discharging experiments. In section IV we introduce our network model and describe the results of the numerical simulations. Finally, we extract some conclusions.

\section{Experimental method}

\subsection{Sample preparation}

\begin{figure}[h]
	\includegraphics[width=.45\textwidth]{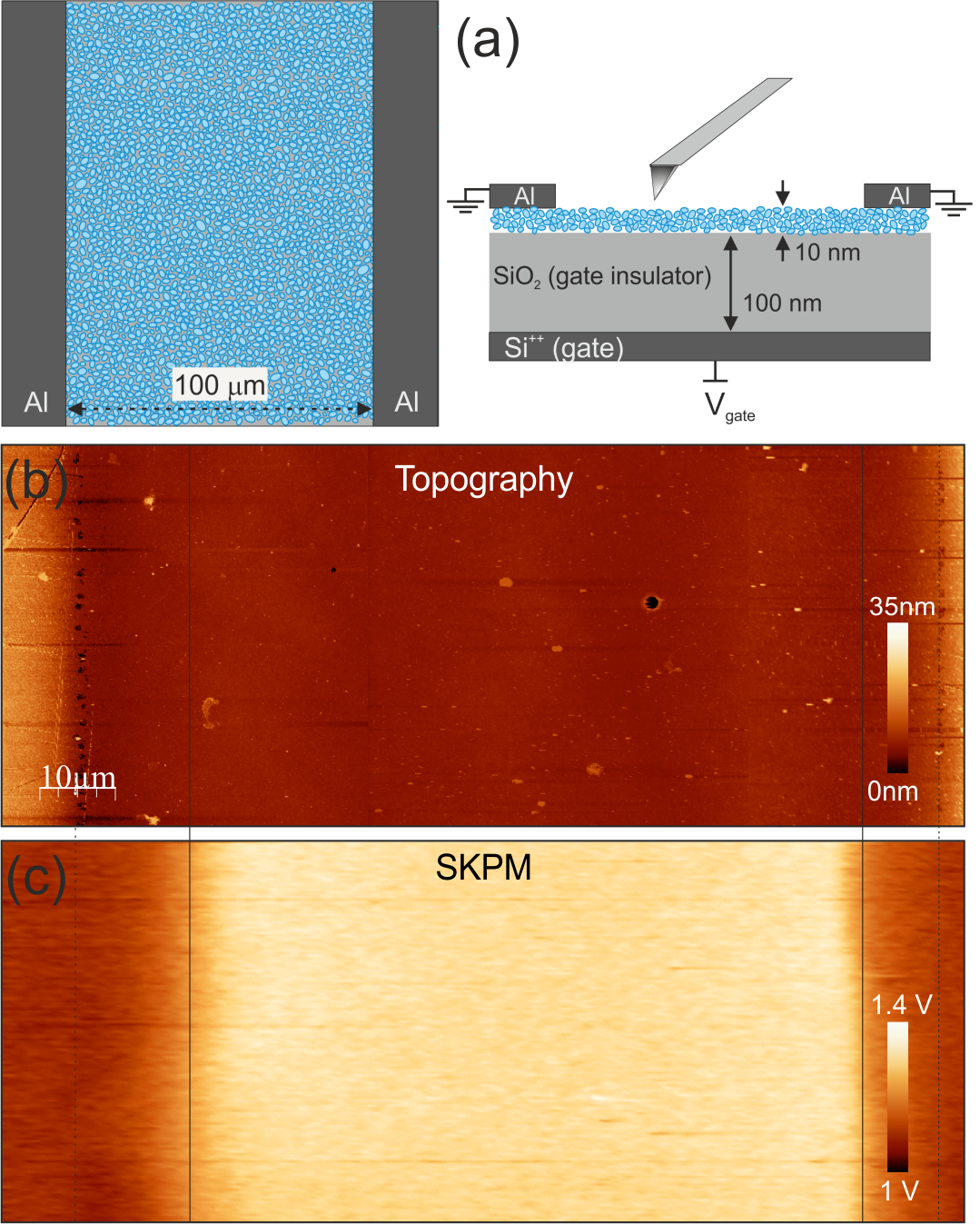}
	\caption{(a)Top and side views of the scheme of the sample. (b)   Topography ($\Delta z= 35$ nm) and (c) SKPM images of the granular aluminum microchannel ($z$ scale 400 mV).}
	\label{fig0}
\end{figure}

The horizontal and vertical schemes of the sample are shown in Fig.\ \ref{fig0} (a).
The granular Al films used in this study were prepared as described elsewhere by the e-beam evaporation of pure Al under a partial pressure of oxygen \cite{Grenet2007}. The parameters here are an Al evaporation rate of 1.8 A/s and a partial oxygen pressure around $3\cdot 10^{-5}$ mbar, for a film thickness of 10 nm. Beyond an oxygen pressure of about $3.2\cdot 10^{-5}$ mbar, the resistance of the films cannot be measured with our experimental set-up at room temperature (their resistance per square is larger than $\approx 100$ G$\Omega$). Previous X rays and TEM investigations have revealed that such films are made of crystalline Al grains of a few nanometers size, dispersed in an amorphous alumina matrix. For the SFM Kelvin probe investigations, the granular Al films were deposited on top of heavily doped Si wafers (the gate) coated with a 100 nm layer of thermally grown SiO$_2$ (the gate insulator). Al contacts 30 nm thick and evaporated on the granular Al film without opening the e-beam evaporation chamber define channels 30-100 micrometers wide (see the topographic picture). No leaking current to the gate was detected (smaller than 1 pA).

\subsection{SFM microscopy}

SFM measurements were carried out using a Nanotec SFM system with a PLL/dynamic measurement board. Once the SFM tip was placed in the granular Al channel between the two lateral contacts  topography and surface potential images were acquired simultaneously by means of  scanning Kelvin probe microscopy (SKPM) at room temperature and ambient conditions using Pt-coated tips (OMCL- AC240TM-R3, nominal $k= 3$ N/m).
To assure quantitative measurements data were acquired in frequency modulation non-contact dynamic mode (FM-DSFM) with an oscillation amplitude of 2 nm while surface potential images were obtained using FM-SKPM mode with $V_{\rm ac} = 500$ mV at 7 kHz. Details of the data acquisition set-up can be found in Ref.\ \onlinecite{palacios09}. Freely available WSxM software has been used for image acquisition and processing \cite{Horcas}.
To improve the time resolution and to minimize the topography crosstalk artifacts \cite{Ortuno2016}, the charging/discharging experiments were performed on a single scanning spatial line, that is, the $y$-scanning was blocked and the $V_{\rm SKPM}$ was measured along the same scanning line. This means that in the corresponding surface potential images the horizontal axis is the position along the line, while the vertical axis is time.

Fig.\ \ref{fig0} (b) shows a topographic image of the granular aluminum microchannel and (c) the SKPM image acquired simultaneously. Since the maximum lateral scanning length of our set-up is $50$x$50$ microns, the panoramic granular Al channel image has been built up from three images taken across the channel, by slightly displacing the tip from image one to another. The images have been acquired with symmetric potentials on source and
drain ($V_{\rm source}=V_{\rm drain}=0$ V) after the gate voltage had been kept at 0 V for three days, so that the sample can be assumed at equilibrium. 
In the topography image (Fig.\ \ref{fig0} (b),
the electric contacts can be recognized as the higher regions ($40\pm 10$ nm) at the left and right 
borders. The lower region corresponds to the very low conductivity granular Al film. 
In the SKPM image the contrast is reversed and those regions have different surface potential, due to their different work function.

\begin{figure*}
	\includegraphics[width=\textwidth]{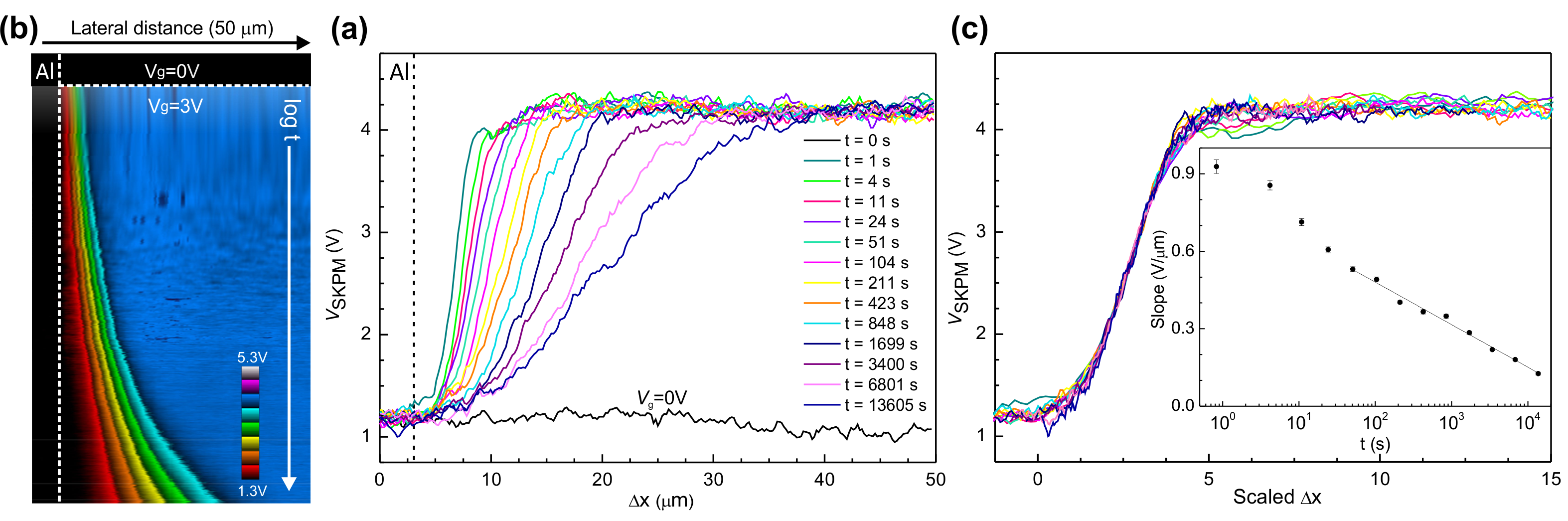}
	\caption{(a) Surface potential image in a region close to a contact (distance in horizontal axis and $\log t$ in the vertical axis). The horizontal dashed line indicates $t=0$ when the gate voltage is switched from 0 V to 3 V. The vertical dashed line marks the position of the effective contact lead. (b) Surface potential profiles from panel (a) at different times. The profile at $V_{\rm g}=0$ has been included as a reference. (c) Same data as in (b) represented versus the scaled lateral distance. Inset: Slope of  $V_{\rm SKPM}$ versus time on a logarithmic scale}
	\label{exp_raw}
\end{figure*}

Before discussing the charging behaviour of the films, we note that
the electric contacts as seen in
the topographic  and   SKPM images do not coincide. In fact, the
low conductivity channel as “seen” by the SKPM image (bright central region, Fig.\ \ref{fig0} (c)) is significantly
narrower than the channel as measured from the topographic image (lower topographic region, Fig.\ \ref{fig0} (b)). As
discussed in more detail elsewhere \cite{palacios05}, the resolution in the FM-
SKPM mode is of the order of 20-50nm, and thus much higher than the discrepancy between the
topographic and SKPM images (of the order of 10 $\mu$m). Instead, we assume that this
discrepancy is due to a shadow effect of the mask used for the evaporation of the gate and drain
contacts: if the mask is not sufficiently close to the Al-grains, some Al may enter the region
between the contacts, increasing the conductivity of the Al-grains near the contacts. Since -within
the height resolution of the topographic images- no step is observed in the lower region, we believe
that most probably this Al diffuses into the granular Al film. In what follows, we will assume that the
“true” Al-grain sample is the region as “seen” in the SKPM image.

\section{Experimental results}

\subsection{Charging experiments}
	
\begin{figure*}
	\includegraphics[width=\textwidth]{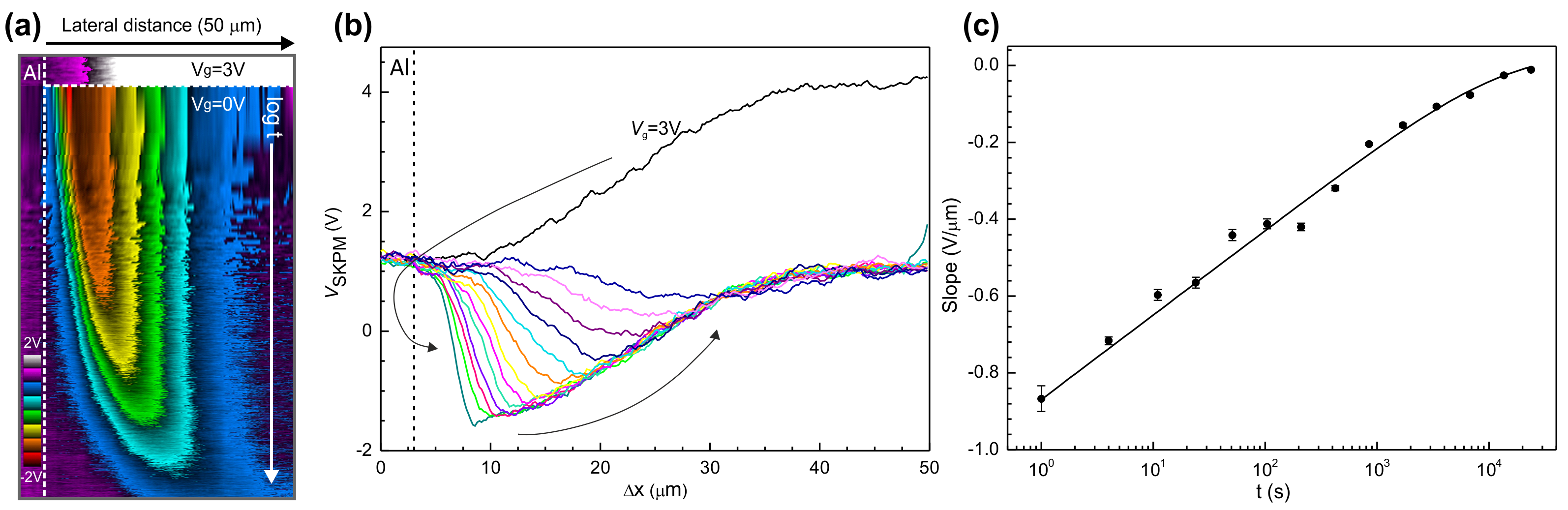}
	\caption{(a) Color map of the data in panel a with distance in horizontal axis and $\log t$ in the vertical axis. (b) Surface potential as a function of distance  at the time intervals exponentially spaced with respect to the instant when  the gate voltage has been switched back to zero. (c) Slopes of the decreasing regions of the curves in panel (b) plotted  versus time. The continuous curve is a fit to Eq.\ (\ref{aging}).}
	\label{exp_rel}
\end{figure*}

To study charging processes in the highly resistive granular Al samples, we first let the system relax to equilibrium for a very long time (typically 3 days) by fixing the two leads and the gate voltage at $V_{\rm g}=0$. 
Then, at some instant that is taken as the time origin, $t=0$, the gate voltage is switched to $V_{\rm g}\neq 0$ and the surface potential $V_{\rm SKPM}$ evolution is monitored. 
Figure\ \ref{exp_raw} (a)  shows  $V_{\rm SKPM}$ as a function of the lateral dimension (horizontal axis) and the time on a logarithmic scale (vertical axis) in a sample region close to one of the leads. Due to scanning size limitations, only one half of the channel is shown. However, it has been checked that a symmetric behavior take places at the other contact.
When  the gate voltage $V_{\rm g}$ is switched from 0 to 3 V, the surface potential of the granular Al suddenly changes by the same amount, while the metal lead potential remains unaffected. 
Just after the voltage switching no current has yet been able to flow into the Al grain film. As a first
approximation, at this instant the sample can therefore be considered as a
dielectric. The SFM tip therefore ``sees'' the gate voltage through the granular Al and the SiO$_2$
insulating layer (see Fig.\ \ref{fig0}).
As time evolves, in order to reach a new equilibrium state,  charge from the source and drain contacts enters into the granular aluminum to screen the $V_{\rm g}$. 
This is seen best in Fig.\ \ref{exp_raw} (b) where $V_{\rm SKPM}$ lateral profiles at several times have been represented. 
These results are fairly independent of the value of the applied gate voltage.
Note that while the time intervals between successive curves grow roughly exponentially, the change in slope between  successive curves shown is similar. 
Thus, we observe an extremely slow  charging behavior, which cannot be explained by a standard diffusive model with low
conductivity.

To quantify the behavior of $V_{\rm SKPM}$  the central region of each curve has been fitted to a straight line
\begin{equation}
V_{\rm SKPM}=a(x-x_0)\;.
\label{fitting}
\end{equation}
All the curves can be overlapped
by plotting them as a function of the scaled variable $\tilde{x}=(x+x_0)/a$ as shown in Figure  \ref{exp_raw} (c). 
We first note the relatively high quality of the overlap. In second place,  the symmetry between the two tails of the curves  is also quite remarkable; we are dealing with an odd function with respect to the  value at the center. 

In the inset of Figure  \ref{exp_raw} (c), we plot the time dependence of the fitted slopes on a semilogarithmic scale. The roughly logarithmic dependence of the slopes with time is noticeable, a fact difficult to understand with uniform models of the conductivity and a clear indication of hopping processes with exponential distributions of resistances.

\subsection{Memory effects}

In our experiments, we have also monitored how the system relaxes back to equilibrium after the gate voltage is switched off.
The results are shown in Fig.\ \ref{exp_rel}a where $V_{\rm SKPM}$ is plotted on a color scale versus the lateral distance (horizontal axis) and time on a logarithmic scale (vertical axis). They are also plotted in Fig\ \ref{exp_rel}b as a function of the lateral distance for several values of the time, with the time intervals between successive profiles increasing exponentially; the origin of time is taken as  the instant in which the gate voltage is switched off.

As in the excitation phase, the curves for different times are roughly equally spaced for the exponentially spaced time intervals chosen. To quantify this fact, we measured the slope of the straight segment before the minimum value of $V_{\rm SKPM}$ for the different curves and the results are plotted versus time on a logarithmic scale in panel {\bf c} of Fig.\ \ref{exp_rel}. Within the experimental error, the data follows a straight line with some bending at long times. The continuous curve corresponds to the standard relaxation behavior observed in systems showing logarithmic time evolutions \cite{Grenet2007,Amir15}
\begin{equation}
x(t)= a \ln \left(1+\frac{t_{\rm w}}{t}\right)\;.
\label{aging}
\end{equation}
where the 'waiting' time $t_{\rm w}$ is the time interval during which the gate has been kept at a certain voltage in the excitation phase (here about  5 h).

\section{Theoretical analysis}

A diffusion model (or, equivalently, a model with homogeneous conductivity) 
cannot explain even qualitatively the main features of these experimental results.
The value of the diffusion constant $D$ establishes the time scale, but cannot get the wide distribution of relaxation times implied by a quasi logarithmic behavior. One can solve the diffusion problem in terms of modes characterized by a wavenumber $k=(\pi n)/L$, where $n$ are integers and $L$ is the distance between the leads. 
The characteristic time of a mode is \cite{adkins}
\begin{equation}
\tau_k=\frac{1}{k^2D}\;,
\label{timemodes}
\end{equation}
resulting in a scaling form of the type $f(x/\sqrt{t})$.
Each mode reduces proportionally at all positions with an exponential time dependence.
The shapes of the potential curves predicted with this type of model are very different from those observed experimentally in the Al-grain samples discussed above. In particular, it is difficult to explain the almost constant value of the potential away from the leads  when its value at closer positions has already changed noticeably.

\subsection{Model}

The logarithmic dependence of the dynamical variables (inset in Fig.\ \ref{exp_raw}c, for example) is indicative of processes with rates depending exponentially on smoothly distributed random variables. 
Based on this hint and on the highly resistive nature of our samples we study a model suitable for hopping processes and adequate for numerical simulations. 
In order to keep the model as simple as possible we construct it in two dimensions.

\begin{figure}[h]
	\includegraphics[width=.35\textwidth]{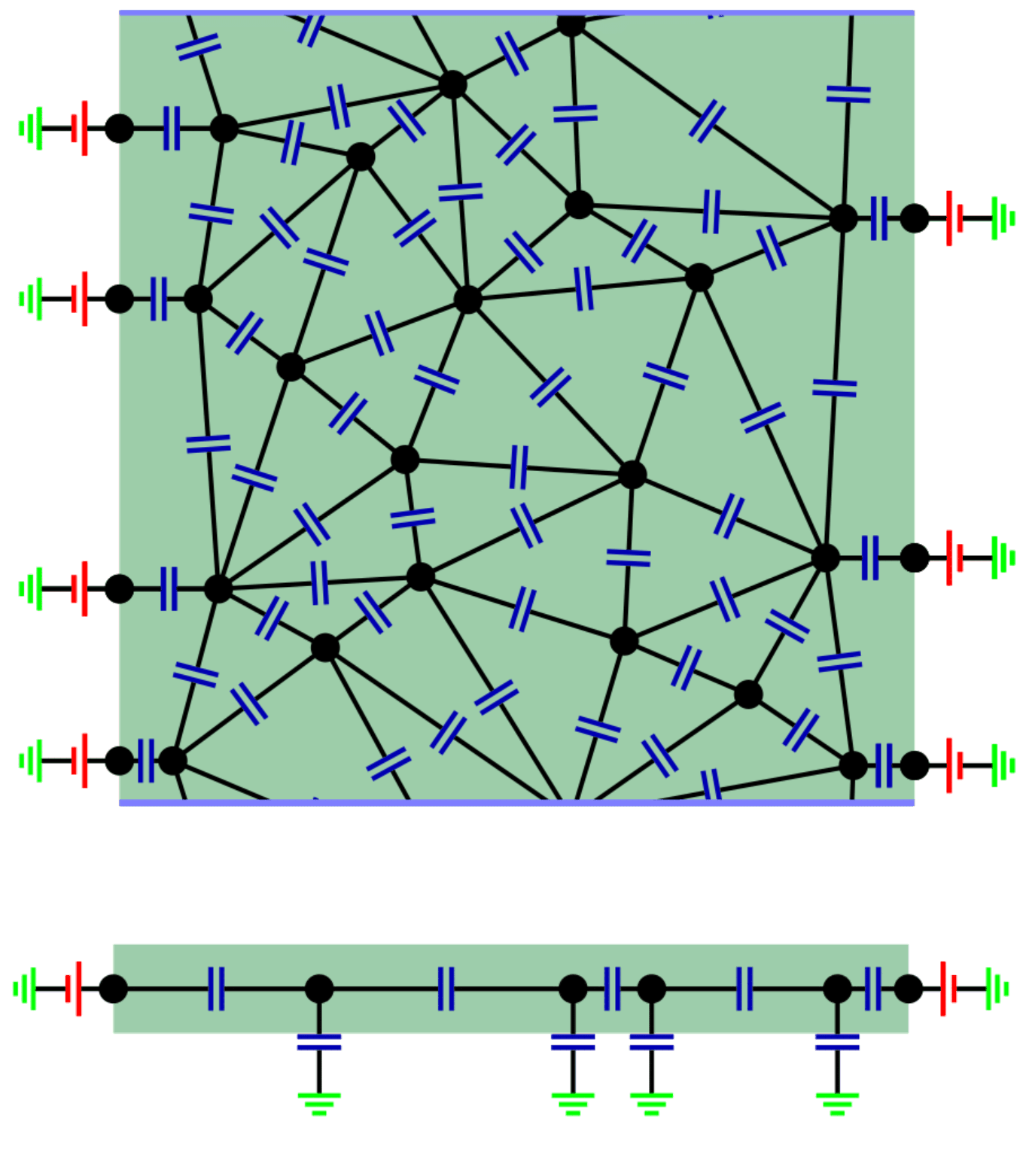}
	\caption{Top and side views of the  scheme  of the capacitors  network used to model the system. Charges can jump between nodes, which is equivalent to having resistances between them (not shown).}
	\label{fig1}
\end{figure}

We consider a network model that  consists on $N$ sites at random on a square sample  of size $L\times L=N$ and capacitors in the links joining nearest neighbor sites.
Each site corresponds to a metallic grain (or set of metallic grains  as explained bellow) and each capacitor to the capacitance between nearest neighbor grains \cite{Ortuno:2015}.
A scheme of the network model is shown in Fig.\ \ref{fig1}.
In order to perform a simulation as close as possible to the experimental setups and at the same time minimize finite size effects, we have included two leads at a potential $V$ in opposite sides of the sample  and periodic boundary conditions in the other two sides. Sites near the leads are connected to them by capacitors. All sites are connected to the gate by capacitances $C_0\epsilon_0$, which take into account the field lines that go out of the system without ending  in a nearby grain.
The capacitances $C_{i,j}$ between sites take random values extracted from the distribution
\begin{equation}
C_{i,j}=C_{\rm r}\epsilon_0 e^{\eta},\qquad \eta\in [-W/2,W/2]\;.
\label{distribution}\end{equation}
We have taken $C_{\rm r}=1$ (which sets our unit of energy) and $W=2$.
The choice of this particular distribution is not crucial, since it just introduces some small randomness in the capacitances and it has the property that the average value of the interaction between nodes  is independent of $W$\cite{Ortuno:2015}. 

The charge $Q_i$ on each site can only take integer values
and is split between the plates of the capacitors connected to this site.  Each capacitor has opposite charges on its two plates
\begin{equation}
Q_i=\sum_j q_{i,j}\qquad q_{i,j}=-q_{j,i}=\sum_{j} C_{i,j} (V_j-V_i)
\end{equation}
where $V_i$ is the potential at site $i$ and the sum on $j$ runs over all sites, the two contacts and the gate.
The charge on the leads can take any (fractional) value.
We can rewrite the previous equation in matrix form
\begin{equation}
Q_i=\sum_j A_{i,j} V_j
\label{new}\end{equation}
where $A_{i,j}$ is the capacitance matrix, defined as
\begin{equation}
A_{i,j}=C_{i,j}-\left(C_0+\sum_{k}C_{i,k}\right)\delta_{i,j} .
\label{elements}\end{equation}
Eq.\ (\ref{new}) can also take care of the leads with a proper extension of the definition
of the charge vector and of the capacitance matrix.
In the leads,
the potential $V_{\rm lead}$ is fixed and the charge is $V_{\rm lead}C_{\rm lead}$, where $C_{\rm lead}$ is the sum of all the capacitances connecting sites of the sample with the lead.
It is convenient to define the following charge vector of dimension $N+2$ (see Ref.\ \cite{Ortuno:2015} for details of the model)
\begin{equation}
{\bf Q}=(Q_1,\cdots,Q_{N},C_{\rm left}V_{\rm left},C_{\rm right}V_{\rm right})
\label{vectorQ}\end{equation}
and to extend the definition of the capacitance matrix to include the conection with the leads.

We usually have an arrangement of charges and want to know the potential
at each site. To solve this problem, we have to invert the capacitance matrix ${\bf F}={\bf A}^{-1}$.
This new matrix ${\bf F}$ plays the role of an effective interaction between charges.
The total energy of the system is directely obatained from  ${\bf F}$
\begin{equation}
H=\frac{1}{2}{\bf Q F Q}^T.
\label{hamil3}\end{equation}

Given a set of charges, we calculate the energy through Eq.\ (\ref{hamil3})
and, following the hopping model of conductivity in localized systems, perform transitions of charge 1 between sites with probability proportional to \cite{POF13}
\begin{equation}
\Gamma_{i,j}=\exp\left(-\frac{2r_{i,j}}{\xi}-\frac{E_{i,j}}{kT}\right)\;.
\label{distance}
\end{equation}
where $r_{i,j}$ is the distance between sites $i$ and $j$, $\xi$  the localization length, $E_{i,j}$ the energy difference due to the transition, $k$ Boltzmann constant and $T$ the temperature. We consider $T=1$ a temperature higher than the average charging energy of a site, but smaller than the applied voltage.
Under these conditions, we expect that the factor of distance in Eq.\ (\ref{distance}) will be the crucial one. The energy factor would at most renormalize the percolation effects produced by the spatial factor, explained below.

The sites in our model can represent real metallic grains or effective ones formed by (possibly) many grains merging together by quantum tunneling in highly connected regions. This will just renormalize the energy scale towards smaller values and the time and spatial scales to larger values.

\subsection{Numerical results}

We have performed Monte Carlo simulations with our network model changing systematically its parameters to try to reproduce as much as possible the experimental results.
The key parameter of the model turns out to be the localization length. When it is larger than the typical site separation, the transition rates between sites are relatively homogeneous and the conduction mechanism can be modeled by diffusion. In the opposite case, hops of charges between close pairs are much more likely than hops between distant sites, the sample is very heterogeneous and conduction can be modeled by percolation theory \cite{POF13}.
As we will see, most experimental features can be explained with the model parameters adjusted to the regime of percolation theory.
 
If the transition rate between two sites is given by Eq.\ (\ref{distance}) and the localization length verifies $\xi\ll 1$ electron jumps between close sites are exponentially more likely that between distant sites. Following the standard approach to percolation in hopping systems \cite{POF13}, we consider that two sites are connected when several hops between them have been produced on average.  Then at  a given time $t$ sites with a separation smaller than
\begin{equation}
r_{i,j}\approx \xi \ln t
\label{separation}
\end{equation}
are connected, while those more separated remain unconnected. 
As explained above, we have assumed that the spatial factor in the hopping rate, Eq.\ (\ref{distance}), is the relevant one for our conditions.
If sites are at random (or grain separations are roughly uniformly distributed), the  exponential dependence of the hopping rate dominates the bond connectivity, and the proportion of bonds connected at a given time  will be of the form
\begin{equation}
p(t)\approx \xi \ln \frac{t}{t_0}
\label{bond}
\end{equation}
where $t_0$ is a constant that ensures the correct initial condition.
Charge can penetrate in the sample along the clusters formed by connected pairs. 
According to percolation theory, the typical size of these clusters $R(t)$ grows with time as \cite{POF13}
\begin{equation}
R(t)\approx \left(p_{\rm c}-p(t)\right)^{-4/3} .
\label{cluster}
\end{equation}
$p_{\rm c}$ is the critical percolation probability for the specific model considered, at which there is an extended connected cluster, and $4/3$ is the universal correlation length exponent for percolation in two dimensions.

\begin{figure}
	\includegraphics[width=\columnwidth]{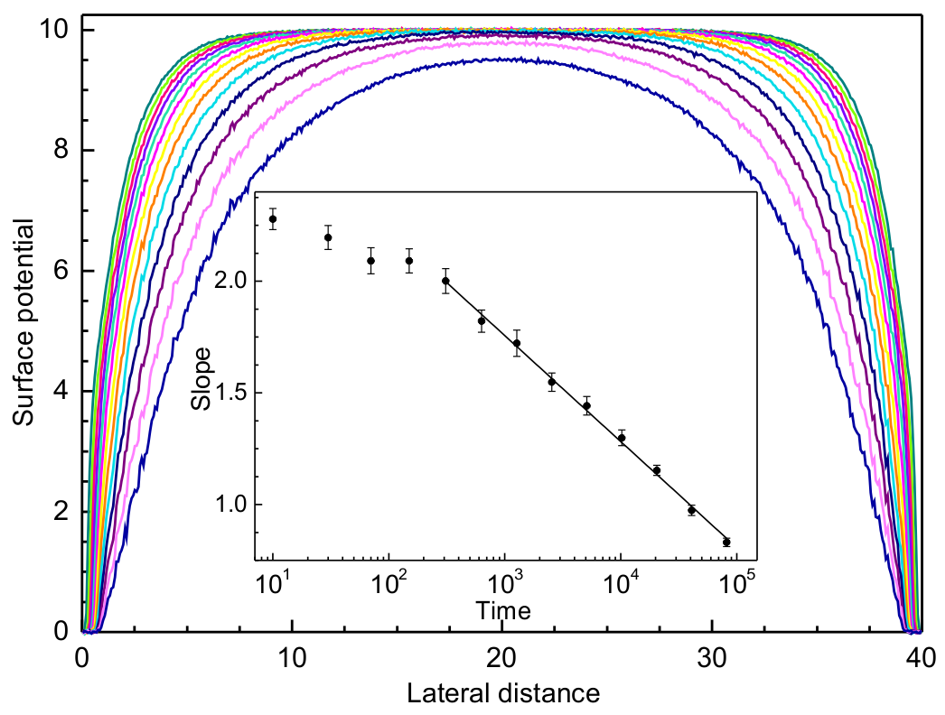}
	\caption{Site potential of our network model as a function of distance for several times exponentially spaced. The localization length is $\xi=0.2$. Inset: Slope of  the site potential at a given distance versus time on a logarithmic scale.}
	\label{teorica}
\end{figure}

As in Kelvin probe microscopy one adjusts the parameters to avoid as much as possible field lines between the point and the surface, the site potential  of our model is directly related to the experimental surface potential. The results of a numerical simulation of the charging process are shown in Fig.\ \ref{teorica}. The average site potential is plotted as a function of the lateral distance to the electrode for several times and measured from the instant in which the gate voltage is switch on. Note that the time interval between successive curves grow exponentially.
The localization length is $\xi=0.2$, so conduction is well in the percolative regime. The overall shape of the curves are similar to the experimental results, but more importantly the time dependence of the slopes of the potential is roughly logarithmic, quite similar to the experimental behavior.
In the inset of Fig.\ \ref{teorica} we represent the slope of the surface potential versus time on a logarithmic scale.
According to our model, the exponentially large distribution of characteristic times is due to the exponential dependence of the hopping rate on distance (and to a roughly uniform distance distribution).
At low temperatures, the energy factor in the hopping rate can contribute further to the broadening of the times involved. As our experiments are at room temperature, we do not expect a significant contribution from the energy factor in the hopping rates.
The gate voltage potential (10 for the results in  Fig.\ \ref{teorica}) is chosen so that the charge in each grain is clearly higher than one. In this case the results are independent of the gate voltage, as in the experiments.

\begin{figure}
	\includegraphics[width=\columnwidth]{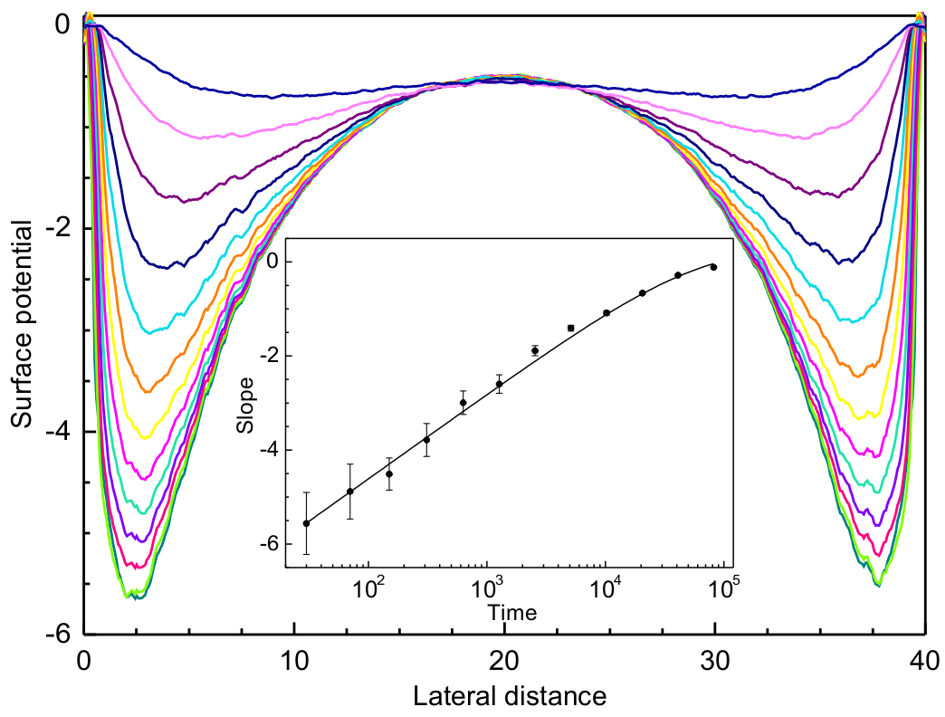}
	\caption{Numerical simulation of the site potential 
 as a function of distance for several times exponentially spaced after the gate voltage has been switch back to zero.  Inset: Slope of the first part of the curves of  the site potential versus time on a logarithmic scale.}
	\label{rel-teorica}
\end{figure}

After the last time for which we measure the charging of the sample, we switch off the gate potential in our simulation and monitor again the time and spatial evolution of the site potential.
The results  are plotted in Fig.\ \ref{rel-teorica} for the same parameters as in the charging process, Fig.\ \ref{teorica}. The main features of the discharging curves are again reproduced by our simulations, although there are differences in the details. The overall logarithmic behavior is well reproduced as can be seen in the inset of Fig.\ \ref{rel-teorica}, where the slopes of the first part of the curves are plotted versus time on a logarithmic scale. The continuous line corresponds to a fit to Eq.\ (\ref{aging}).

\begin{figure}
	\includegraphics[width=\columnwidth]{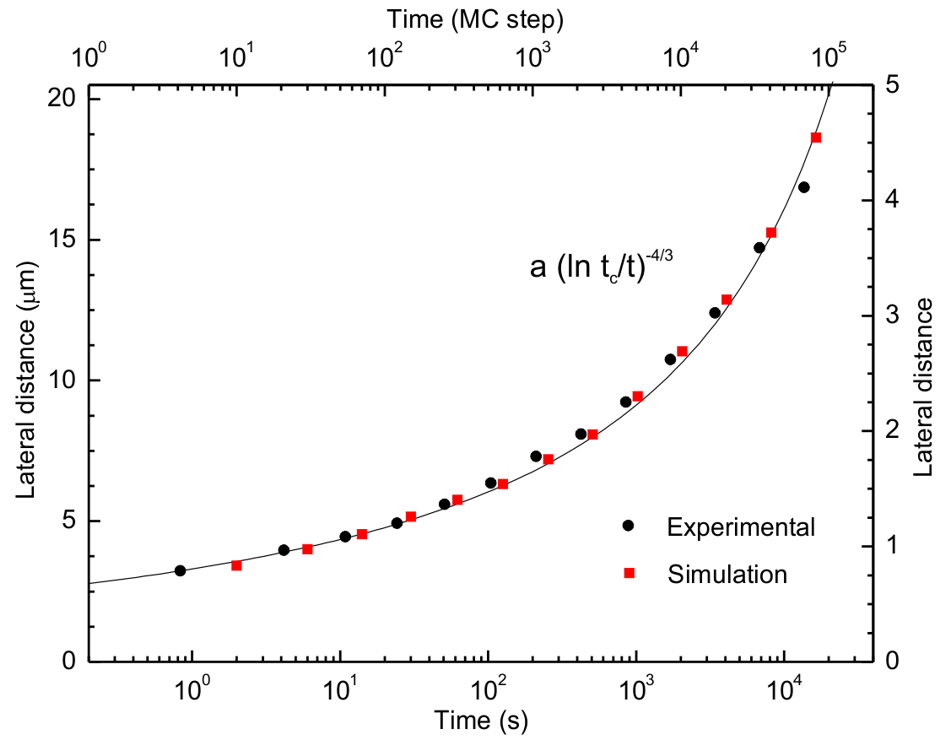}
	\caption{Lateral distance of the point where the surface potential is equal to the mean value between the potential in the lead and in the middle of the sample as a function of time for the experimental results shown in Fig.\ \ref{exp_raw} (black points, bottom and left axis) and for the simulations (red points, top and right axis).}
	\label{perco}
\end{figure}

We can have a more restrictive test of our percolation model.
If one pays attention to the experimental curves, Fig.\ \ref{exp_raw}, one can observe that they seem to accelerate (in a logarithmic scale) their approach to equilibrium at large time intervals. This is clearly evident in panel {\bf a} where the initially almost straight lines become bent at the bottom of the figure.
We interpret this as a sign that, at the largest times involved in our experiments, we are close to reaching percolation, so that the size of the connected clusters increases faster, according to Eq.\ \ref{cluster}, as the system approaches critical percolation and the whole sample would be connected.
To quantify this observation we plot the lateral distance of the point at which the voltage is equal to the mean between the lead and the gate voltages as a function of time.
In Fig.\ \ref{perco}, we show the results for both the experimental data (black points, left axis in $\mu$m) and the numerical simulations (red points, right axis in grains separation). The horizontal time axis is in seconds for the experimental points and in Monte Carlo sweeps for the simulation results, and it has been shifted for the simulation results to achieve a maximum overlap between both set of points.
The line is a fit of the experimental data to the analytical expression of the typical cluster size $R$ in percolation theory
\begin{equation}
R(t)= a \left(\ln \frac{t_{\rm c}}{t}\right)^{-4/3}\;.
\label{cluster}
\end{equation}
To derive this equation, we have taken into account the time dependence of the connecting probability, Eq.\ (\ref{bond}), $t_{\rm c}$ is the critical time needed to reach percolation, and $4/3$ is the correlation exponent in two-dimensional percolation.
The quality of the fit and the analogy between experimental and numerical simulation results is strong indication of the percolative character of the charging process.

\section{Conclusions}

We have repeated the previous measurements for other samples with different resistances. If the samples are conducting enough so that its conductivity can be measured, the processes studied here take place at a fast scale and our set up is not fast enough to detect intermediate processes. If the samples are much more resistive than the ones analyzed here, we do not see any dynamics. As resistance varies exponentially with grain separation, we can only measure a relatively narrow range of metallic concentrations.

In principle, this technique can be adapted to measure the conductivity of highly resistive samples for which standard macroscopic methods are not sensitive enough to determine it. We now intend to apply the SKPM technique to study at the nanometer scale the glassy effects reported in the conductance of several disordered insulators.\cite{Grenet2007,Ovadyahu2007,HaEi12, Grenet2010}. Indeed in some of them, in the appropriate resistance range, these occur up to room temperature.\cite{Eisenbach2016, Delahaye2014, granular} The technique can be very useful to analyze memory effects, and combined with conductance measurements, may shine light on the mechanisms involved.

	\section*{Acknowledgements}
	The work was supported by the
	Spanish MINECO Grants No. FIS2015-67844-R, ENE2013-48816-C5-1-R, ENE2016-79282-C5-4-R, and CSD2010-00024, and by Fundaci\'on S\'eneca grant  19907/GERM/15.


\end{document}